\documentclass{WileyMSP-template}
\usepackage{amsmath,amssymb,amsthm,amsfonts}
\usepackage[numbers,sort&compress]{natbib}
\usepackage[labelfont=bf]{caption}

\begin{document}

\title{Advancing Plasmonic Computing with Single-Beam Logic Primitives}

\maketitle


\author{Komal Gupta}
\author{Anand Hegde}
\author{Chen-Bin Huang*}


\dedication{}

\begin{affiliations}
Komal Gupta, Anand Hegde, and Prof. Chen-Bin Huang*\\
Institute of Photonics Technologies, National Tsing Hua University, Hsinchu 30013, Taiwan \\
Email: robin@ee.nthu.edu.tw\\

\end{affiliations}



\begin{abstract}
\justifying
Plasmonic logic circuits combine ultrafast operation with nanoscale integration, making them a strong candidate for next-generation optical computing. Realizing this potential, however, requires overcoming practical challenges such as bulky interferometric designs and reliance on secondary control signals. This work advances plasmonic logic by introducing a single-global threshold mechanism in plasmonic two-wire transmission lines, empowered with polarization modal selectivity and geometric tuning to enable versatile circuit functionality. The scheme embeds the control signal with a single laser beam, supporting six deterministic polarization states and eliminating the need for auxiliary inputs. With this framework, we experimentally demonstrate advanced logic operations, including a 2-bit comparator, parity checkers, and encoder/decoder circuits. The approach reduces circuit footprint by 67\% and power consumption by 50\% relative to state-of-the-art systems, while maintaining low latency and high stability. By unifying thresholding, polarization, and geometry into a compact, source-free scheme, this work pushes plasmonic nanocircuitry from device-level novelty toward scalable, energy-efficient architectures for next-generation optical processors.

\end{abstract}


\section{Introduction}
\justifying
Optical computing has gained serious traction in recent years, driven by advances in photonic integration, low-loss waveguides, and a surge of application interest in optical neural networks and accelerators \cite{markov2014limits,li2021challenges}. Its appeal lies in inherent parallelism, extreme bandwidth, and the potential for low-energy logic operations that sidestep the thermal bottlenecks of traditional CMOS \cite{agrell2016roadmap}. Yet many of the hybrid systems face scalability hurdles. Alternative approaches such as carbon-nanotube logic and biochemical circuits offer conceptual versatility but fall short on throughput and reproducibility \cite{bachtold2001logic,elbaz2010dna}. Even in plasmonic systems, most existing logic gates rely on interferometric structures or wavelength filtering, which demand extra optical sources or control beams and complicate large-scale integration \cite{wei2011cascaded,peng2018universal,lu2014chip}. These challenges highlight the gap between current device-level demonstrations and scalable architectures.

Despite these obstacles, plasmonic circuitry remains one of the most promising paths forward. Its ability to confine light below the diffraction limit while preserving ultrafast propagation enables dense layouts and high-speed logic, capabilities critical for scaling optical processors beyond what free-space or multi-beam architectures can support \cite{ozbay2006plasmonics}. The experimental realization of Surface Plasmon Polaritons (SPPs) interference-based logic gates and circuits has been achieved through diverse design strategies to demonstrate logical operations \cite{drezet2007plasmonic,zhao2010plasmonic,wei2012nanowire,fu2012all}.  These designs enable precise routing of SPPs and manipulation of the desired near-field intensity distribution by orchestrating interference effects among SPPs traversing nanowire networks \cite{PhysRevApplied.1.014007,razinskas2016normal}. While the near-field signal provides comprehensive information, detection of output signals at the far-field predominantly relies on distinguishing between strong and weak intensities, corresponding to `1' and `0' states, respectively, through the establishment of an output threshold. 

Essential logic circuits often invert outputs producing a ``1" when both inputs are ``0" which appears to violate energy conservation, necessitating an external control signal to resolve the paradox \cite{peng2018universal,fu2012all,maram2020frequency,chang2023enhancing}. Semiconductor electronics use added voltage sources, while all-optical logic gates rely on secondary lasers, both increasing system complexity and energy demands. Current plasmonic approaches face a core limitation in their dependence on these external controls, especially for logic functions that require high outputs from all-zero inputs. Overcoming this limitation without introducing further complexity would significantly improve the scalability, reproducibility, and integration potential of plasmonic logic circuits.

We address this challenge with a single-beam plasmonic logic framework built on the two-wire transmission line (TWTL), which supports symmetric and antisymmetric modes selectable by polarization and geometry \cite{PhysRevApplied.1.014007,razinskas2016normal,geisler2013multimode,dai2014mode,chen2020polarization}. Control is encoded internally through six polarization states from one laser, eliminating auxiliary sources, while a single-global thresholding (SGT) scheme digitally enforces logic states including the critical (0,0→1) transition without added mechanisms. Modal selectivity replaces phase-sensitive interferometry, enabling robust traveling-wave operations \cite{wu2022broadband}. Within this architecture, we realize compact logic blocks such as a 2-bit comparator, parity checkers, and encoder/decoder units, reducing circuit footprint by up to 67\% and power consumption by 50\% with a single source \cite{wei2011cascaded,peng2018universal,wei2012nanowire,fu2012all,chang2023enhancing,wei2011quantum,wang2016nanoscale}. These gains mitigate intrinsic losses and layout variability while supporting dense, reproducible designs. For optical and hybrid accelerators, where bandwidth and density must scale without control overhead, collapsing the control path into the data beam and using the TWTL’s modal basis as the logic alphabet eliminates chronic blockers: extra lasers and phase-sensitive optics while preserving plasmonics’ core strengths of confinement and bandwidth. The result is circuit-level functionality that tiles, composes, and reconfigures, positioning single-beam, polarization-controlled primitives as a practical route to compact, low-power plasmonic logic fabrics aligned with multidimensional photonic-computing roadmaps \cite{bente2025potential, wu2025field}.

\section{Design and Working Principle}
\justifying
We implement single-beam logic primitives on a plasmonic TWTL using dimer and link input antennas with a single-stub mode detector for readout \cite{dai2014mode}. Only the outer scatter of the stub is taken as the logic output and the inner spot is ignored. As minimal ingredients, we use the known polarization responses of the two couplers: for the dimer, left-circular and right-circular polarizations (LHCP/RHCP) are routed into branches with asymmetric intensities \cite{chen2020polarization,wu2022broadband}; linear $0^{\circ}/90^{\circ}$ states excite two equal-intensity branch termini (low and high at our operating points, respectively) \cite{geisler2013multimode}; and linear $\pm45^{\circ}$ states produce routing with branch intensity asymmetry. The link (not shown here) provides a complementary routing palette referenced previously \cite{wu2022broadband}.

The transition from antenna behavior to a quantitative description is captured by a matrix map that incorporates antenna-specific geometry. With input polarization $(\theta,\delta)$, the initial complex field amplitudes on branches $A$ and $B$ are
\begin{equation}
\label{eq:pol_coupling}
\begin{bmatrix}
w_A(0)\\[2pt] w_B(0)
\end{bmatrix}
=\,\mathbf{M}^{(\mathrm{type})}_{\mathrm{ant}}(a,b)\,
\begin{bmatrix}
\cos\theta\\ e^{i\delta}\sin\theta
\end{bmatrix}E_0,\qquad \mathrm{type}\in\{\mathrm{dimer},\mathrm{link}\},
\end{equation}
where $\mathbf{M}_{\mathrm{ant}}$ encodes the antenna geometry (e.g., dimer and link length $a$, width $b$) and any deliberate branch-amplitude asymmetry, $w_{A,B}(0)$ denote the complex field amplitudes launched on branches $A$ and $B$, and $E_0$ is the incident field amplitude. The six programmed states are:
\begin{equation*}
    (\theta,\delta)\in\{(0^{\circ},0),\, (90^{\circ},0),\, (+45^{\circ},0),\, (-45^{\circ},0),\, (45^{\circ},+\pi/2),\, (45^{\circ},-\pi/2)\}
\end{equation*}
which together span linear and circular excitations.

The two branch signals propagate and coherently combine at the outer port of the single-stub. Branch-amplitude asymmetry (set by $\mathbf{M}_{\mathrm{ant}}$) provides a first degree of freedom to tailor the output intensity, a geometry-tunable phase (bends/asymmetry) provides a second, enabling enforcement of the desired superposition:
\begin{equation}
\label{eq:outer_superposition}
I_{\mathrm{out}}
=\Bigl|\,c_A\,w_A(0)\,e^{i\phi_A}+c_B\,w_B(0)\,e^{i(\phi_B+\phi_{\mathrm{geom}})}\Bigr|^2,
\end{equation}
where $c_{A,B}$ capture the branch-to-stub projection, $\phi_{A,B}$ are the accumulated phases along the two paths, and $\phi_{\mathrm{geom}}$ is a geometry-controlled phase offset.

Logic is assigned with one threshold $I_{\mathrm{th}}$ applied identically at input branch ends and at the output:
\begin{equation}
A=[I_A(0)\ge I_{\mathrm{th}}],\quad
B=[I_B(0)\ge I_{\mathrm{th}}],\quad
\mathrm{Output}=[I_{\mathrm{out}}> I_{\mathrm{th}}]
\end{equation}
where each bracketed term evaluates to $1$ if the condition is true and $0$ otherwise. This SGT criterion is a concise replacement for dual-thresholding: it directly ties input classification to output digitization with a single beam and no auxiliary signals.

The design principles in selected cases for equal branch lengths with $\phi_{\mathrm{geom}}=0$ ( \textbf{Figure 1}a) are: for $0^{\circ}$ on a dimer, both branch termini are equal-intensity and sub-threshold (inputs $A=B=0$), and by the intrinsic symmetry of the TWTL for dimer coupling at the outer port in Equation~\ref{eq:outer_superposition} adds constructively, yielding $I_{\mathrm{out}}>I_{\mathrm{th}}$ and the key outcome $(0,0)\!\rightarrow\!1$. For $90^{\circ}$ on a dimer, both branch termini are equal-intensity and above threshold (inputs $A=B=1$) while the outer port is suppressed by the antisymmetric projection, giving Out $=0$. For $\pm45^{\circ}$ on a dimer, routing with branch intensity asymmetry produces $(1,0)$ or $(0,1)$ at the inputs, the superposition is imperfect and weak, yielding $I_{\mathrm{out}}<I_{\mathrm{th}}$. For LHCP/RHCP on the dimer, polarization routing yields the largest branch-intensity contrast (one branch dominant), after superposition at the outer port, $I_{\mathrm{out}}>I_{\mathrm{th}}$ (Out $=1$). Across these six programmed polarizations, $\phi_{\mathrm{geom}}$ serves as a final knob to enforce the required superposition while keeping a single, global threshold. The details of antenna parameters and their optimization are described in Supporting Information S2.\\

To empirically validate the SGT criterion’s robustness, we fabricate a half-circuit comparator (dimer with two open branches, no mode detector) and read the terminal scattering as $A$ and $B$. Figure~1b shows a scanning electron microscope (SEM) of the gold TWTL comparator. The fabrication and measurement procedures using a home-built confocal microscopy setup are described in the Experimental Section and in Supporting Information S1, respectively. Figure~1c reports the experimentally demonstrated comparator states using a normalized intensity threshold of $I_{\mathrm{th}}=0.7$. The panel comprises four sub-images arranged in a $2\times2$ layout: (top-left) excitation with $0^{\circ}$ yields $A=B=0$, and the comparator output is $1$, (top-right) RHCP realizes the $A>B$ case, (bottom-left) LHCP realizes the $B>A$ case both giving the comparator output $0$, (bottom-right) excitation with $90^{\circ}$ yields $A=B=1$, and the comparator output is $1$. The branch-terminus intensity ratios underpin the state assignment in all four cases, and applying the same $I_{\mathrm{th}}$ formalizes classification consistently for subsequent circuits.

By unifying polarization programming, geometry-enabled superposition (via $\mathbf{M}_{\mathrm{ant}}$ and $\phi_{\mathrm{geom}}$), and a single global threshold, this realization advances plasmonic computing toward deterministic logic in a single-beam, subwavelength footprint, providing a scalable foundation for the circuit blocks that follow. \\

We next realize parity checkers using symmetric branches and the same single-stub mode detector. Antenna parameters and the six-state polarization responses are summarized in the Supporting Information (Section~S2). \textbf{Figure~2 (a and c)} show SEM image of the Even/Odd parity checker devices based on the dimer and link antennas, respectively. Figure~2 (b and d) compare experimental scatter maps (top rows) with the corresponding FDTD simulations (bottom rows), enabling one-to-one verification under the SGT.

 The dimer-based device in Figure~2 (a and b), produces the largest branch-intensity contrasts for LHCP and RHCP excitations by routing SPPs into a single branch alternatively giving $(0,1)\!\rightarrow\!1$ and $(1,0)\!\rightarrow\!1$ cases, respectively. In each of the above cases, the designated projection coefficients $c_{A,B}$ and the geometry-controlled phase $\phi_{\mathrm{geom}}$ determine the superposition. Even though possessing contrasting amplitudes, the fields carried by branches add predominantly in phase to produce $I_{\mathrm{out}}>I_{\mathrm{th}}$. The parity function with output $Y=0$ is implemented  for the equal-high states through perpendicular polarization whereas no excitation represents the equal-low state. The $(1,1)\!\rightarrow\!0$ case uses the $90^{\circ}$ state, giving equal, above-threshold terminal intensities but an antisymmetric combination at the detector, the outer-port projection then enforces near-$\pi$ phase opposition between the branch contributions, resulting in destructive interference and $I_{\mathrm{out}}<I_{\mathrm{th}}$.

The link-based device in Figure~2(c and d) realizes the complementary parity. With $0^{\circ}$ excitation, both termini are sub-threshold yet phase-aligned through the link’s coupling map, so the outer port receives a constructive sum and $(0,0)\!\rightarrow\!1$. For $(1,1)\!\rightarrow\!1$, either LHCP or RHCP jointly brightens both branches while preserving an in-phase projection at the detector, giving a strong outer-port intensity above threshold. Linear $\pm45^{\circ}$ states route asymmetrically and bias one branch, the resultant amplitude imbalance, combined with the detector’s projection, reduces the net field at the outer port so that $(1,0)\!\rightarrow\!0$ and $(0,1)\!\rightarrow\!0$. Across Figure~2(b and d), each case exhibit the same interference-driven trends-constructive superposition for the accepted parity states and weak or canceling superposition for the rejected states-consistent with Equation~\ref{eq:outer_superposition} and the SGT.\\

\textbf{Figure~3a} shows the 4-to-2 Encoder: a device comprised of two three-branch units with inputs $(Y_3,Y_2,Y_1,Y_0)$ and outputs $(A_1,A_0)$.  This approach is inspired by past research where similar 
circuits were employed to achieve plasmonic half-subtractors and demultiplexers \cite{wu2022broadband}. The input coupler is polarization-programmable (design rules and near-field evolution is discussed in Supporting Information Section~S3). The geometry imposes fixed phase relations at the readout stubs by introducing path-length offsets: wire lengths and bends are chosen so that $\Delta\phi(Y_0,Y_3)\!\approx\!\pi$, while the third branch is approximately in phase with $Y_3$ at the detector. At each port, the detected field is the coherent sum of the three traveling SPPs, whether the phasors in Equation~\ref{eq:outer_superposition} add or cancel is therefore set by these path-engineered phases.

Figure~3b comprises four columns (left to right: $-45^{\circ}$, $90^{\circ}$, $0^{\circ}$, $+45^{\circ}$), with experiment in the top row and corresponding simulations in the bottom row, classification uses a normalized threshold $I_{\mathrm{th}}=0.55$. For the four one-hot inputs defined by the programmed polarization states, superposition at the two outputs proceeds as follows. 
$-45^{\circ}$: power is routed to $Y_0$ and arranged to be anti-phase with its companion branch at both ports, so both sums cancel and $(A_1,A_0)=(0,0)$, i.e., $(0,0,0,1)\!\rightarrow\!(0,0)$. 
$90^{\circ}$: phases are engineered to add at $A_0$ and oppose at $A_1$, yielding $(0,1)$, i.e., $(0,0,1,0)\!\rightarrow\!(0,1)$. 
$0^{\circ}$: the relative phase is flipped, producing addition at $A_1$ and cancellation at $A_0$, giving $(1,0)$, i.e., $(0,1,0,0)\!\rightarrow\!(1,0)$. 
$+45^{\circ}$: routing favors $Y_3$ together with its in-phase companion, so both ports add constructively and $(A_1,A_0)=(1,1)$, i.e., $(1,0,0,0)\!\rightarrow\!(1,1)$. 
These outcomes follow directly from the path-set phases and the coherent-sum rule in Equation~\ref{eq:outer_superposition}.\\

\textbf{Figure~4a} presents the SEM image of the 2-to-4 decoder, which reverses the encoder logic by mapping two inputs $(A_1,A_0)$ to four outputs $(Y_3,Y_2,Y_1,Y_0)$. Unlike the encoder, which uses a compact three-branch design to produce binary output via path-length-based phase cancellation, the decoder fans out from a central input to four arms, each routed to a distinct output port. Each output arm corresponds to a specific output bit in the one-hot decoding logic. The functional principle remains the same: traveling SPPs launched into the arms interfere coherently at each output, and constructive or destructive interference determines the digital state via the same global threshold criterion. The branch geometry, phase offsets, and coupling profiles are engineered to ensure that each polarization state selectively brightens a single output while suppressing the others.

Figure~4b shows the decoder operation across four programmed polarization states, with experimental maps in the top row and corresponding FDTD simulations in the bottom row. The classification threshold is $I_{\mathrm{th}}=0.6$, applied identically at all ports. For $0^\circ$ input, the symmetric mode is excited. Due to the geometry and balanced arm lengths, constructive interference is designed to occur only at $Y_0$, giving the state $(A_1,A_0) = (0,0) \rightarrow (0,0,0,1)$. All other outputs remain below threshold, confirmed by the isolated bright spot in both experiment and simulation. Under $-45^\circ$ polarization, asymmetric coupling launches power into the appropriate branches that produce constructive interference only at $Y_1$. This gives $(A_1,A_0) = (0,1) \rightarrow (0,0,1,0)$. The branching asymmetry is engineered so that the destructive interference dominates at $Y_0$, $Y_2$, and $Y_3$. With $+45^\circ$ polarization, the routing flips compared to the $-45^\circ$ case. Constructive interference now occurs at $Y_2$, yielding $(A_1,A_0) = (1,0) \rightarrow (0,1,0,0)$. The orthogonal polarization ensures phase alignment is reversed, and the branch design compensates accordingly. At $90^\circ$, the antisymmetric mode dominates, and the geometry enforces constructive interference at $Y_3$, giving $(A_1,A_0) = (1,1) \rightarrow (1,0,0,0)$. The symmetric cancellation in the other arms leaves them below threshold.

Design considerations are detailed in the Supporting Information (Section~S4). The decoder uses a mixture of dimer and link antennas, with path-length asymmetry and modal projection tailored to each output bit. Compared to the encoder, which manipulates a minimal phase basis to compress logic, the decoder expands the input state into a spatial distribution by engineering distinct constructive interference conditions at each output. Both encoder and decoder thus leverage traveling-wave superposition in TWTL networks, governed by polarization-programmed input and passive phase-engineered geometry.

\section{Conclusion}
This work demonstrates a single-beam plasmonic logic framework that addresses one of the most persistent barriers in optical nanocircuitry: the reliance on auxiliary inputs and phase-sensitive architectures. By integrating a single-global threshold mechanism into plasmonic two-wire transmission lines and harnessing polarization-selective modes with geometric tuning, we collapse control and computation into a single laser beam. This enables deterministic logic states, including the critical (0,0→1) transition, while supporting advanced operations such as comparators, parity checkers, and encoder/decoder circuits. The approach achieves up to 67\% footprint reduction \cite{wei2011cascaded,wei2012nanowire, wei2011quantum} and 50\% lower power consumption compared to cascaded designs, without compromising speed or stability \cite{fu2012all,peng2018universal,chang2023enhancing,wang2016nanoscale,sang2018broadband}.

The broader significance lies in shifting plasmonic logic from isolated demonstrations toward scalable, energy-efficient circuit fabrics. Encoding control internally through polarization transforms plasmonic nanocircuits into reproducible, densely integrated units that mitigate losses and variability while retaining plasmonics’ unmatched confinement and bandwidth. This positions polarization-controlled plasmonic primitives as a practical platform for high-density optical processors, hybrid photonic–electronic accelerators, and future multidimensional computing systems. Looking forward, expanding logic complexity, refining polarization encoding, and merging with silicon photonics could propel this framework toward functional, wafer-scale architectures. By unifying simplicity and scalability, this work opens a pathway to plasmonic logic as a cornerstone of next-generation information technologies.

\section{Experimental Section}

The optical measurements were performed using home-built confocal microscopy setup (discussed in detail in supporting information section S1).

\threesubsection{Numerical Simulations} All the numerical simulations were performed using Ansys Lumerical FDTD. The device geometry was designed using the optical constants of gold from Johnson and Christy’s material model \cite{johnson1972optical}. The circuitries were excited using a 1560nm femtosecond laser source. 

\threesubsection{Fabrications} Thin Au-film was deposited on glass substrate. The devices were fabricated using a Focused Ion Beam (FIB) system. During the fabrication, the ion beam current was maintained between 7.2pA and 7.5pA to mill the devices. The surrounding regions were etched using the ion beam current of 24pA.\\

\medskip
\textbf{Supporting Information} \par 
Supporting Information is available from the Wiley Online Library or from the author.

\medskip
\textbf{Acknowledgements}\par
The authors gratefully acknowledge Ting-Yin Chen for contributing to the conceptual development of the En/De-coder working principle. \\
This work was supported by the National Science and Technology Council, Taiwan under Grant 112-2112-M-007-021-MY3.

\medskip


\bibliographystyle{MSP}
\bibliography{reference}

\begin{thebibliography}{10}
\providecommand{\url}[1]{\texttt{#1}}
\providecommand{\urlprefix}{URL }

\bibitem{markov2014limits}
I.~L. Markov,
\newblock \emph{Nature} \textbf{2014}, \emph{512}, 7513 147.

\bibitem{li2021challenges}
C.~Li, X.~Zhang, J.~Li, T.~Fang, X.~Dong,
\newblock \emph{PhotoniX} \textbf{2021}, \emph{2}, 1 20.

\bibitem{agrell2016roadmap}
E.~Agrell, M.~Karlsson, A.~Chraplyvy, D.~J. Richardson, P.~M. Krummrich,
  P.~Winzer, K.~Roberts, J.~K. Fischer, S.~J. Savory, B.~J. Eggleton, et~al.,
\newblock \emph{Journal of optics} \textbf{2016}, \emph{18}, 6 063002.

\bibitem{bachtold2001logic}
A.~Bachtold, P.~Hadley, T.~Nakanishi, C.~Dekker,
\newblock \emph{Science} \textbf{2001}, \emph{294}, 5545 1317.

\bibitem{elbaz2010dna}
J.~Elbaz, O.~Lioubashevski, F.~Wang, F.~Remacle, R.~D. Levine, I.~Willner,
\newblock \emph{Nature nanotechnology} \textbf{2010}, \emph{5}, 6 417.

\bibitem{wei2011cascaded}
H.~Wei, Z.~Wang, X.~Tian, M.~K{\"a}ll, H.~Xu,
\newblock \emph{Nature communications} \textbf{2011}, \emph{2}, 1 387.

\bibitem{peng2018universal}
C.~Peng, J.~Li, H.~Liao, Z.~Li, C.~Sun, J.~Chen, Q.~Gong,
\newblock \emph{ACS photonics} \textbf{2018}, \emph{5}, 3 1137.

\bibitem{lu2014chip}
C.~Lu, X.~Hu, H.~Yang, Q.~Gong,
\newblock \emph{Scientific reports} \textbf{2014}, \emph{4}, 1 3869.

\bibitem{ozbay2006plasmonics}
E.~Ozbay,
\newblock \emph{science} \textbf{2006}, \emph{311}, 5758 189.

\bibitem{drezet2007plasmonic}
A.~Drezet, D.~Koller, A.~Hohenau, A.~Leitner, F.~R. Aussenegg, J.~R. Krenn,
\newblock \emph{Nano letters} \textbf{2007}, \emph{7}, 6 1697.

\bibitem{zhao2010plasmonic}
C.~Zhao, J.~Zhang,
\newblock \emph{Acs Nano} \textbf{2010}, \emph{4}, 11 6433.

\bibitem{wei2012nanowire}
H.~Wei, H.~Xu,
\newblock \emph{Nanophotonics} \textbf{2012}, \emph{1}, 2 155.

\bibitem{fu2012all}
Y.~Fu, X.~Hu, C.~Lu, S.~Yue, H.~Yang, Q.~Gong,
\newblock \emph{Nano letters} \textbf{2012}, \emph{12}, 11 5784.

\bibitem{PhysRevApplied.1.014007}
C.~Rewitz, G.~Razinskas, P.~Geisler, E.~Krauss, S.~Goetz, M.~Paw\l{}owska,
  B.~Hecht, T.~Brixner,
\newblock \emph{Phys. Rev. Appl.} \textbf{2014}, \emph{1} 014007.

\bibitem{razinskas2016normal}
G.~Razinskas, D.~Kilbane, P.~Melchior, P.~Geisler, E.~Krauss, S.~Mathias,
  B.~Hecht, M.~Aeschlimann,
\newblock \emph{Nano letters} \textbf{2016}, \emph{16}, 11 6832.

\bibitem{maram2020frequency}
R.~Maram, J.~v. Howe, D.~Kong, F.~D. Ros, P.~Guan, M.~Galili, R.~Morandotti,
  L.~K. Oxenl{\o}we, J.~Aza{\~n}a,
\newblock \emph{Nature Communications} \textbf{2020}, \emph{11}, 1 5839.

\bibitem{chang2023enhancing}
K.-H. Chang, Z.-H. Lin, P.-T. Lee, J.-S. Huang,
\newblock \emph{Scientific reports} \textbf{2023}, \emph{13}, 1 5020.

\bibitem{geisler2013multimode}
P.~Geisler, G.~Razinskas, E.~Krauss, X.-F. Wu, C.~Rewitz, P.~Tuchscherer,
  S.~Goetz, C.-B. Huang, T.~Brixner, B.~Hecht,
\newblock \emph{Physical review letters} \textbf{2013}, \emph{111}, 18 183901.

\bibitem{dai2014mode}
W.-H. Dai, F.-C. Lin, C.-B. Huang, J.-S. Huang,
\newblock \emph{Nano letters} \textbf{2014}, \emph{14}, 7 3881.

\bibitem{chen2020polarization}
T.-Y. Chen, D.~Tyagi, Y.-C. Chang, C.-B. Huang,
\newblock \emph{Nano Letters} \textbf{2020}, \emph{20}, 10 7543.

\bibitem{wu2022broadband}
P.-Y. Wu, Y.-C. Chang, C.-B. Huang,
\newblock \emph{Nanophotonics} \textbf{2022}, \emph{11}, 16 3623.

\bibitem{wei2011quantum}
H.~Wei, Z.~Li, X.~Tian, Z.~Wang, F.~Cong, N.~Liu, S.~Zhang, P.~Nordlander,
  N.~J. Halas, H.~Xu,
\newblock \emph{Nano letters} \textbf{2011}, \emph{11}, 2 471.

\bibitem{wang2016nanoscale}
F.~Wang, Z.~Gong, X.~Hu, X.~Yang, H.~Yang, Q.~Gong,
\newblock \emph{Scientific reports} \textbf{2016}, \emph{6}, 1 24433.

\bibitem{bente2025potential}
I.~Bente, S.~Taheriniya, F.~Lenzini, F.~Br{\"u}ckerhoff-Pl{\"u}ckelmann,
  M.~Kues, H.~Bhaskaran, C.~D. Wright, W.~Pernice,
\newblock \emph{Nature Reviews Physics} \textbf{2025}, 1--12.

\bibitem{wu2025field}
T.~Wu, Y.~Li, L.~Ge, L.~Feng,
\newblock \emph{Nature Photonics} \textbf{2025}, 1--8.

\bibitem{sang2018broadband}
Y.~Sang, X.~Wu, S.~S. Raja, C.-Y. Wang, H.~Li, Y.~Ding, D.~Liu, J.~Zhou,
  H.~Ahn, S.~Gwo, et~al.,
\newblock \emph{Advanced Optical Materials} \textbf{2018}, \emph{6}, 13
  1701368.

\bibitem{johnson1972optical}
P.~B. Johnson, R.-W. Christy,
\newblock \emph{Physical review B} \textbf{1972}, \emph{6}, 12 4370.

\end{thebibliography}


\section*{Figures}
\begin{figure}
\centering
  \includegraphics[width=0.7\linewidth]{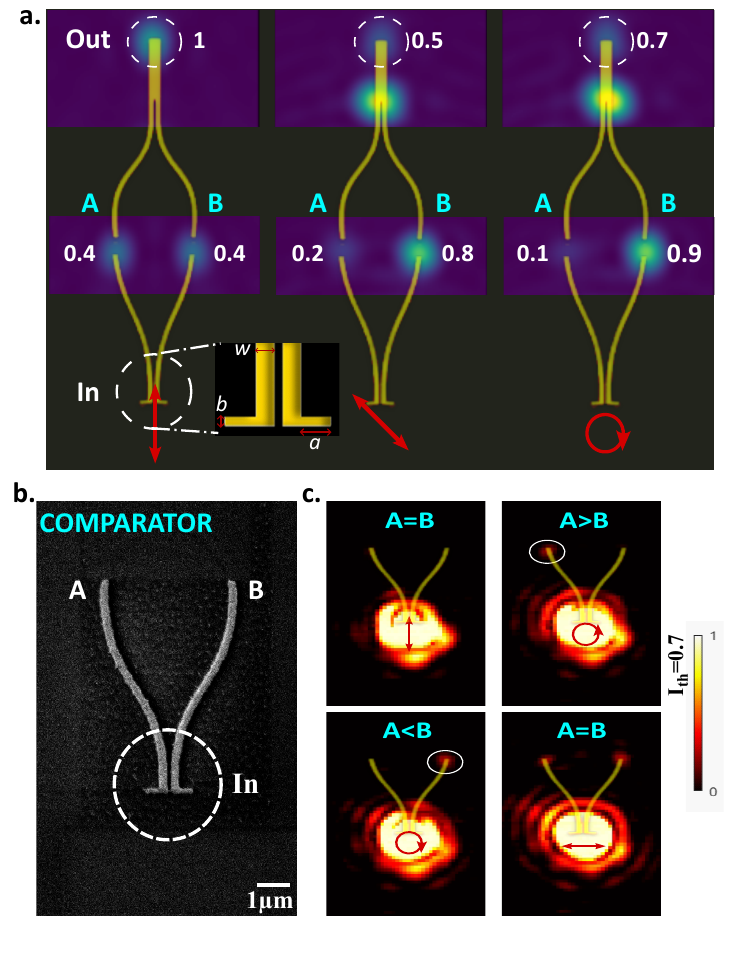}
  \caption{ Single-beam TWTL primitive and comparator establishing the single-global threshold (SGT) criterion. 
(a) Working principle illustrated with simulated intensity maps. A dimer input routes polarization-programmed SPPs into branches $A$ and $B$. The single-stub mode detector is read only at the outer scatter (Out). Numbers indicate normalized intensities at the branch termini and at Out (dashed circle). Equal, sub-threshold inputs ($A\!=\!B$) add constructively at the outer port to realize $(0,0)\!\rightarrow\!1$; asymmetric routing (linear $\pm45^{\circ}$) yields an imbalanced pair with a weak superposition at the outer port $(1,0)/(0,1)\!\rightarrow\!0$; circular polarization produces the largest branch-contrast and a strong constructive sum, giving Out $=1$. Inset shows dimer antenna parameters $(w,a,b)$ used in the simulations. 
(b) SEM image of the gold half-circuit comparator (two open branches, no mode detector).
(c) Experimental validation of input classification with corresponding polarization labels and the same normalized threshold ($I_{\mathrm{th}}=0.7$). The four panels show $A\!=\!B$ (0$^{\circ}$, both sub-threshold), $A\!>\!B$ (RHCP), $A\!<\!B$ (LHCP), and $A\!=\!B$ (90$^{\circ}$, both above threshold). The terminal-intensity ratios reproduce the intended input states and anchor the SGT used for all subsequent circuits.}
  \label{fig:boat1}
\end{figure}

\begin{figure}
\centering
  \includegraphics[width=0.7\linewidth]{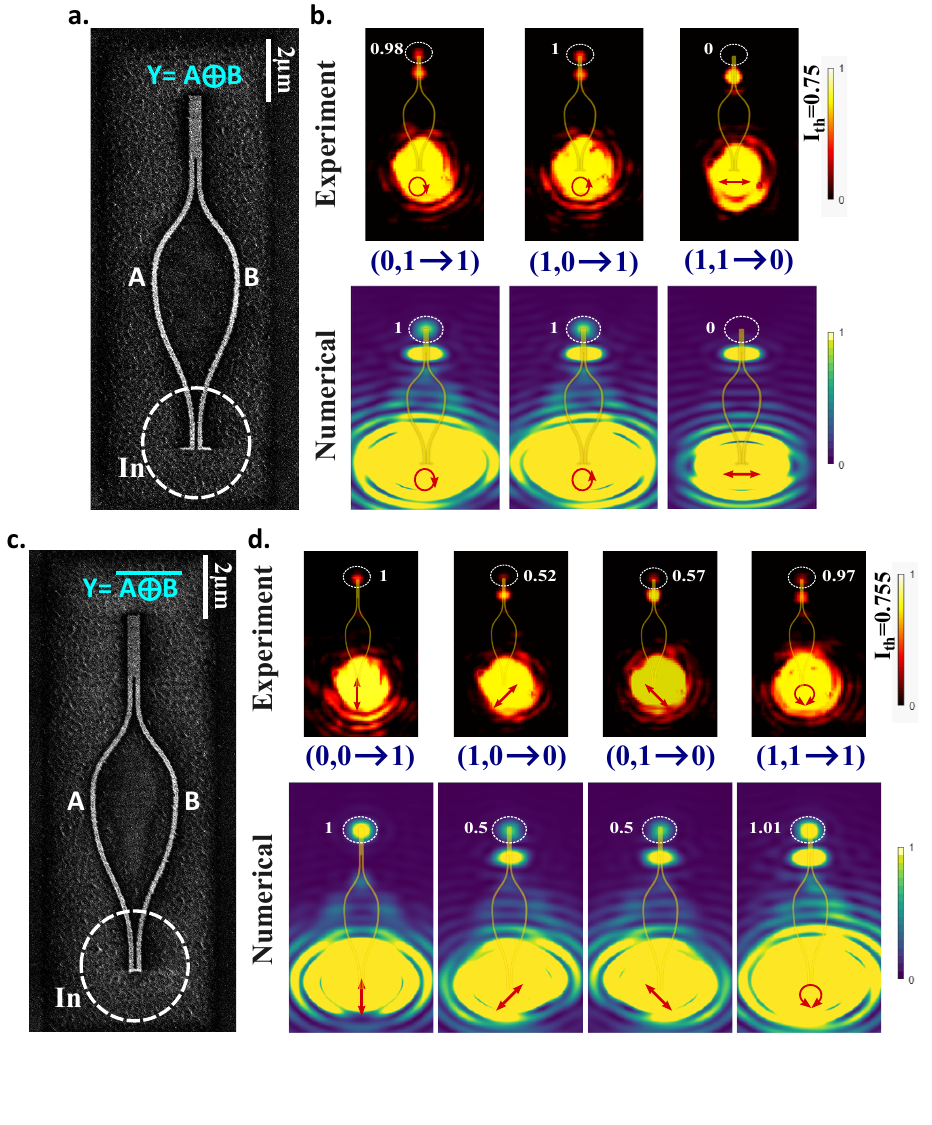}
  \caption{ Parity checkers. (a) SEM image of the dimer-based even-parity device. (b) Output maps with $I_{\mathrm{th}}=0.75$. The top panel shows experiment and the bottom panel shows numerical results. LHCP and RHCP route SPPs into a single branch which yields $(0,1\!\to\!1)$ and $(1,0\!\to\!1)$ at the readout, the $90^{\circ}$ state equalizes branch amplitudes and projects anti-symmetrically to give $(1,1\!\to\!0)$. (c) SEM image of the link-based odd-parity device. (d) Output maps with $I_{\mathrm{th}}=0.755$. The top panel shows experiment and the bottom panel shows numerical simulations. The $0^{\circ}$ state keeps both termini low yet phase aligned so the outer port adds to $(0,0\!\to\!1)$, LHCP or RHCP brighten both branches in phase to give $(1,1\!\to\!1)$, linear $\pm45^{\circ}$ bias one branch and weaken the projected field to give $(1,0\!\to\!0)$ and $(0,1\!\to\!0)$.}
\end{figure}

\begin{figure}
\centering
  \includegraphics[width=\linewidth]{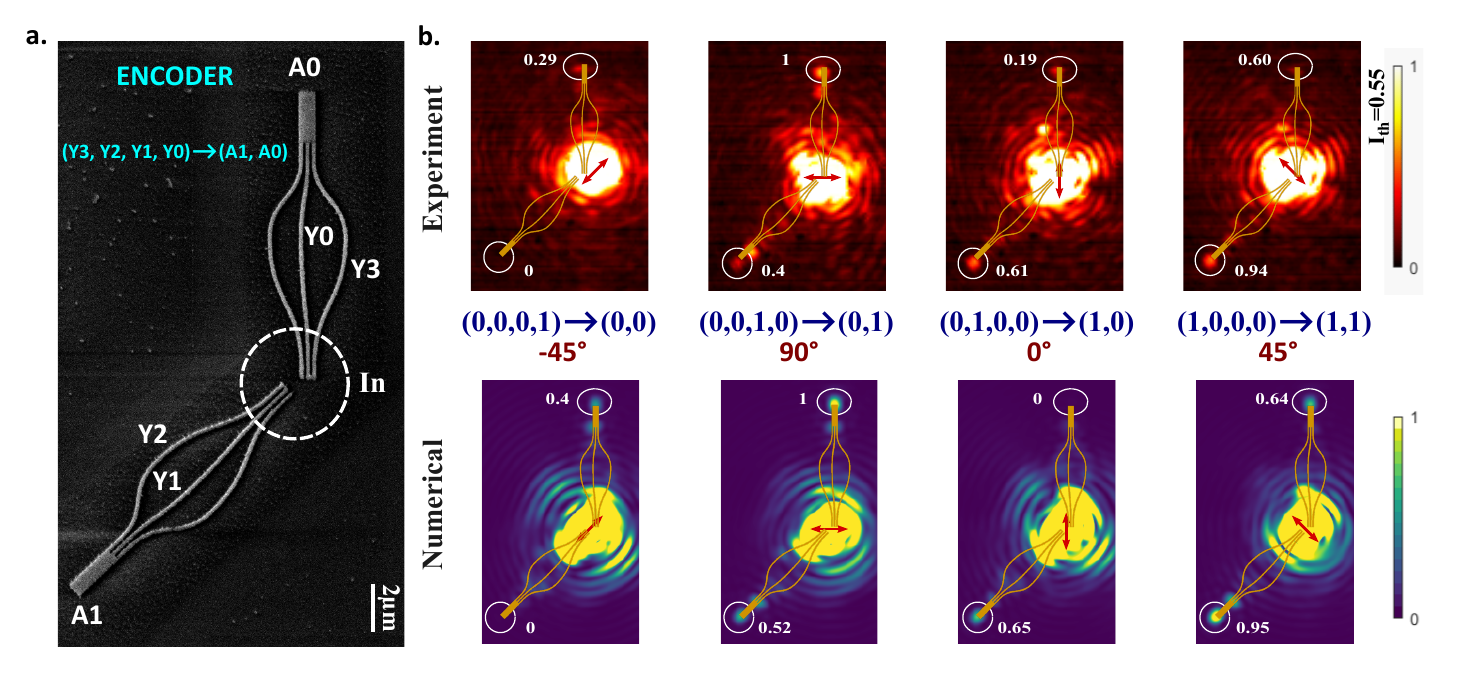}
  \caption{ Encoder. (a) SEM image of the 4–to-2 encoder built from two three-branch units. Path-length offsets impose $\Delta\phi(Y_{0},Y_{3})\approx\pi$ and set the companion branch of $Y_{3}$ in phase at the readout to program addition or cancellation. (b) Output maps with $I_{\mathrm{th}}=0.55$. The top panel shows experiment and the bottom panel shows numerical results. Columns from left to right use input polarizations $-45^{\circ}$, $90^{\circ}$, $0^{\circ}$, and $+45^{\circ}$. For $-45^{\circ}$ the routing selects $Y_{0}$ and both ports are arranged anti-phase so $(0,0,0,1)\!\rightarrow\!(0,0)$. For $90^{\circ}$ the fields add at $A_{0}$ and oppose at $A_{1}$ so $(0,0,1,0)\!\rightarrow\!(0,1)$. For $0^{\circ}$ the relative phase is flipped so $(0,1,0,0)\!\rightarrow\!(1,0)$. For $+45^{\circ}$ the routing favors $Y_{3}$ with an in-phase companion so both ports add and $(1,0,0,0)\!\rightarrow\!(1,1)$. Coherent superposition at each port follows the phase-engineered paths and yields the desired input-output combinations.}
  \label{fig:boat1}
\end{figure}

\begin{figure}
\centering
  \includegraphics[width=\linewidth]{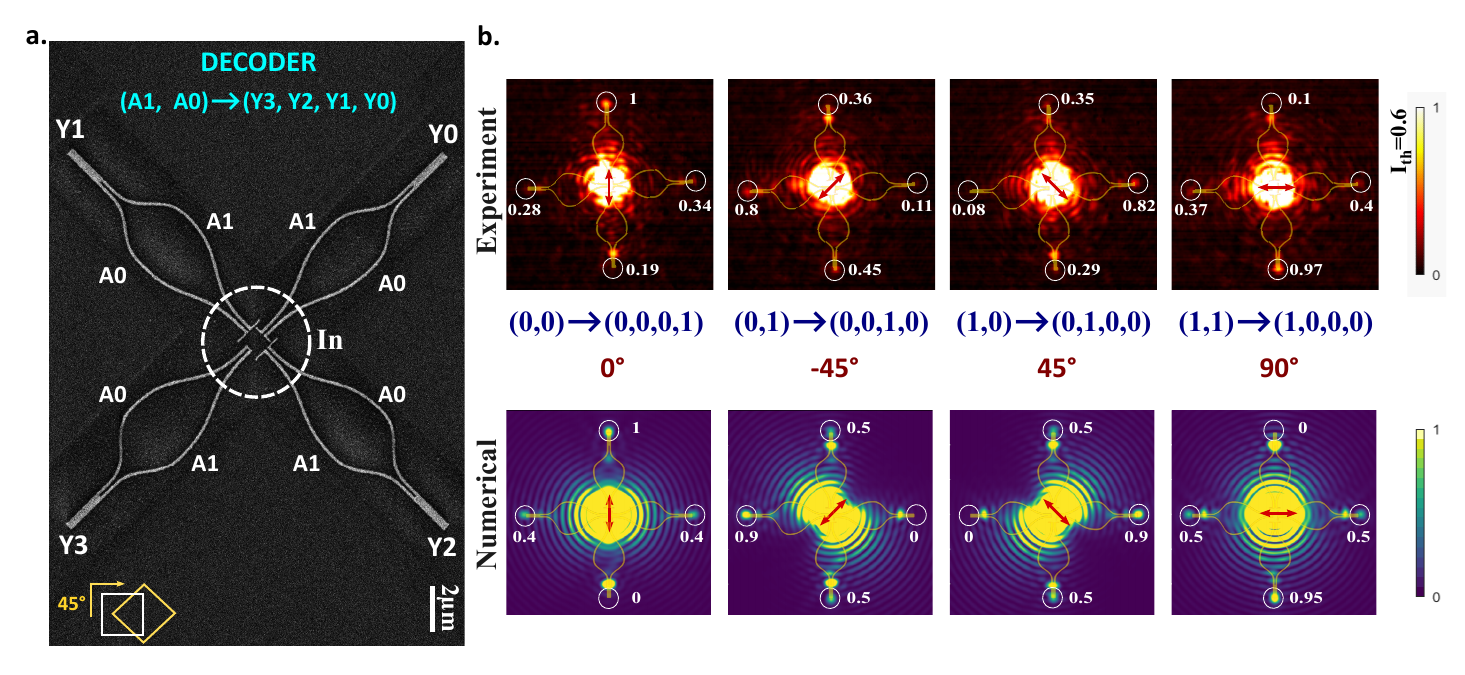}
  \caption{Decoder. (a) SEM image of the 2-to–4 decoder that maps inputs $(A_{1},A_{0})$ to one-hot outputs $(Y_{3},Y_{2},Y_{1},Y_{0})$ by path-engineered phase and modal projection at each readout. (b) Output maps with $I_{\mathrm{th}}=0.6$. The top panel shows experiment and the bottom panel shows numerical results. Columns use input polarizations $0^{\circ}$, $-45^{\circ}$, $+45^{\circ}$, and $90^{\circ}$. For $0^{\circ}$ constructive interference is designed only at $Y_{0}$ so $(0,0)\!\rightarrow\!(0,0,0,1)$. For $-45^{\circ}$ the routing and phase set addition at $Y_{1}$ so $(0,1)\!\rightarrow\!(0,0,1,0)$. For $+45^{\circ}$ constructive interference occurs at $Y_{2}$ so $(1,0)\!\rightarrow\!(0,1,0,0)$. For $90^{\circ}$ the antisymmetric mode brightens $Y_{3}$ so $(1,1)\!\rightarrow\!(1,0,0,0)$. The same threshold is applied at all ports and the observed intensities match the designed superposition.}
  \label{fig:boat1}
\end{figure}


\begin{figure}
\textbf{\\In this work, plasmonic two-wire transmission lines are advanced with a single-global threshold mechanism that eliminates auxiliary inputs while supporting multiple deterministic states through polarization control and geometric tuning. A 2-bit comparator, parity checkers, and encoder/decoder are experimentally demonstrated. This approach advances nanoelectronics and paves the way for compact, high-performance optical computing devices, showcasing the forefront of plasmonic-based technologies for future information processing systems.}\\

\medskip
\centering
  \includegraphics{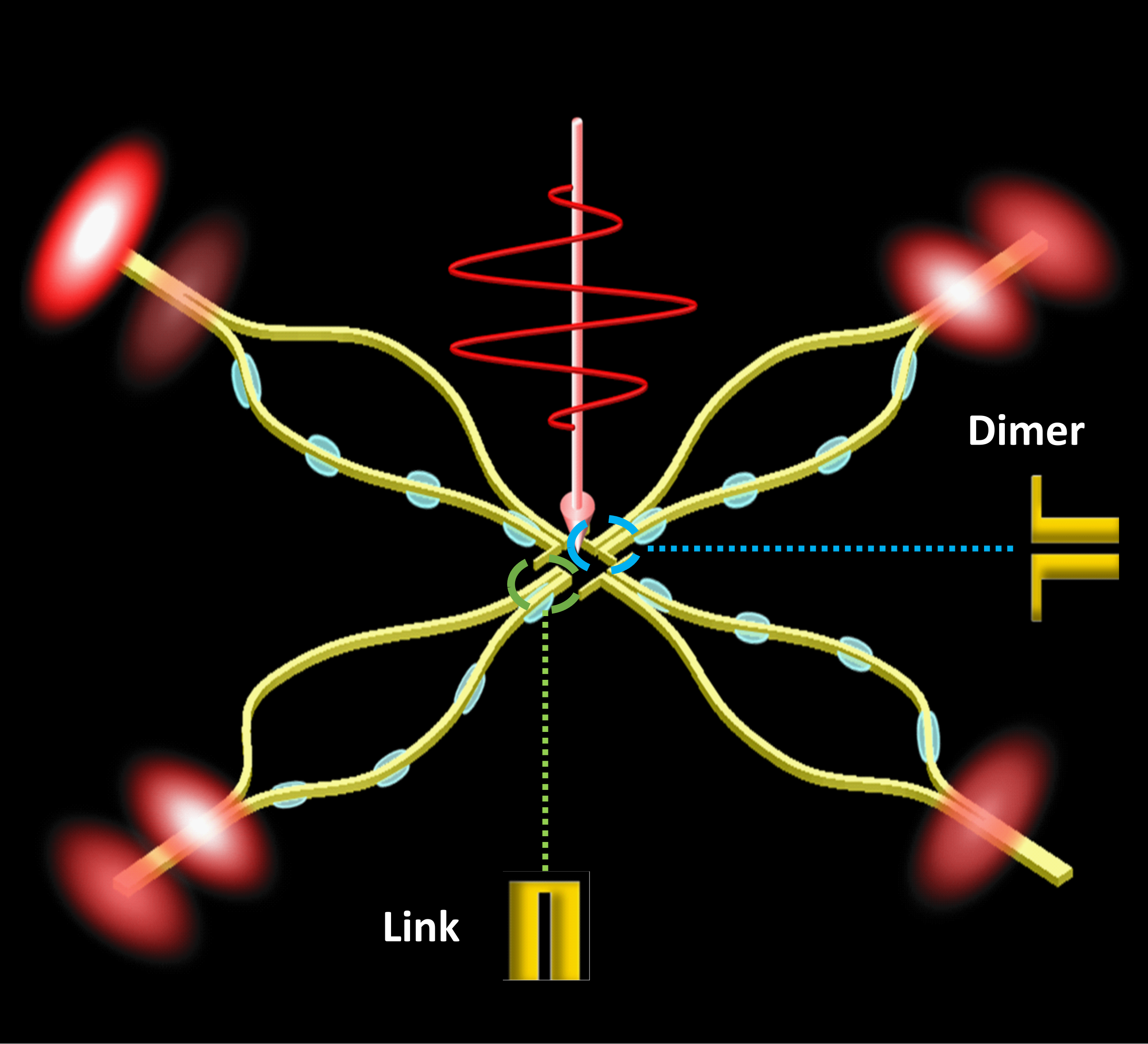}
  \medskip
  \caption*{Advancing Plasmonic Computing with Single-Beam Logic Primitives}
\end{figure}

\end{document}